\documentclass[aps,twocolumn,groupedaddress,showpacs,reprint]{revtex4-2}
\usepackage{amsmath}
\usepackage{bm}

\begin{document}

\title{
Beyond low-dissipation regime: general condition for Curzon-Ahlborn efficiency
}

\author{Yuki Izumida}
\affiliation{Department of Complexity Science and Engineering, Graduate School of Frontier Sciences, The University of Tokyo, Kashiwa 277-8561, Japan}
\thanks{izumida@k.u-tokyo.ac.jp}

\begin{abstract}
In recent years, physicists have extensively investigated the universality of Curzon-Ahlborn (CA) efficiency as the efficiency at maximum power of heat engines, 
which, like Carnot efficiency, is determined only by heat-bath temperatures.
While the universality of the CA efficiency to the first order in the temperature difference is established within linear response theory in the inverse operation time, 
it is generally lost at quadratic order.
In the low-dissipation Carnot heat engine model known as the standard framework, the symmetric dissipation condition is necessary. Yet, it is not a necessary condition as the original CA model that adopts endoreversibility does not require such symmetry. 
In this study, we extend the low-dissipation model to include nonlinear response in the inverse operation time and derive a general condition under which the CA efficiency is recovered in the quadratic order. 
The generalized symmetric dissipation condition and the CA model with Newton's law of cooling are shown to be special cases of this condition.
By applying the obtained condition, we also clarify the condition under which an endoreversible model with a general heat-transfer law exhibits the CA efficiency.
\end{abstract}

\maketitle

Carnot's discovery of the maximum efficiency of heat engines in the nineteenth century~\cite{C1824} laid the groundwork for the birth of thermodynamics.
Even today, Carnot's insight is still essential for stochastic~\cite{PE1996,MRDPPR2016,VEWV2014} or quantum heat engines~\cite{QLSN2007,KL2014,VA2016} at the micro- and nano-scale, opening new frontiers in thermodynamics across scales.

For any heat engine operating between hot and cold heat baths at temperatures $T_h$ and $T_c$, its maximum efficiency is given by the Carnot efficiency $\eta_{\rm C}$:
\begin{align}
\eta_{\rm C}\equiv 1-\frac{T_c}{T_h}.\label{eq.Carnot}
\end{align}
Because achieving the Carnot efficiency generally requires quasistatic operation, the power, the work output per unit time, vanishes.
This motivates the study of heat engines that deliver finite, and preferably maximum, power.
Beyond its practical importance, this consideration led to another fundamental result concerning the efficiency of heat engines.

More than half a century ago, Curzon and Ahlborn in their celebrated paper~\cite{CA1975} derived the efficiency at maximum power $\eta^*$ of a simple model of a Carnot heat engine operating in finite time (see also Refs.~\cite{Y1955,C1957,N1958}):
\begin{align}
\eta^*=1-\sqrt{\frac{T_c}{T_h}}\equiv \eta_{\rm CA} \ \ ({\rm CA efficiency}).\label{eq.CA}
\end{align}
The remarkable feature of this result is its simplicity and universality, which, like the Carnot efficiency (\ref{eq.Carnot}), depends only on the heat-bath temperatures.
Their derivation was phenomenological and assumed endoreversibility, under which irreversibility arises only from heat transfer between the working substance and the heat baths, and a specific heat transfer law~\cite{R1979,C1985}.
The discovery of the CA efficiency later gave birth to finite-time thermodynamics beyond the problem of heat engines~\cite{ASB1984,ABOS1984,AS2026}.
See also Refs.~\cite{VLFR2014,MP2015,AC2025} for historical perspectives on the formula (\ref{eq.CA}).

In recent years, the universality of the CA efficiency has been extensively investigated by physicists beyond endoreversible assumption, heat transfer laws, and scales~\cite{CVB2005,JCH2007,BSC2011,SS2008,ELV2009,EKLV2010,IO2012,IO2014,ST2014,ST2015,WT2012,WT2012_2,YG2012}.
Expanding $\eta_{\rm CA}$ in powers of $\eta_{\rm C}$ gives
\begin{align}
\eta_{\rm CA}=\frac{\eta_{\mathrm C}}{2}+\frac{\eta_{\mathrm C}^2}{8}+O(\eta_{\rm C}^3).
\end{align}
The linear term of $\eta^*$ is determined by the leading linear response in the inverse operation time, and the universality of the CA coefficient $1/2$ is well established~\cite{CVB2005,SS2008,EKLV2010,IO2014}.
In contrast, the quadratic order of $\eta^*$ can differ from the CA coefficient $1/8$, showing that the CA efficiency is no longer universal in this order~\cite{SS2008,ELV2009,EKLV2010,IO2012,ST2014,ST2015,WT2012,WT2012_2,YG2012}.
In the low-dissipation (LD) Carnot heat engine model~\cite{EKLV2010}, which considers the linear response with respect to the inverse operation time, the CA efficiency in the quadratic order or even in full order is recovered under the symmetric dissipation condition. 
The LD model provides a standard framework to describe finite-time heat engines, which applies to a broad class of models without specifying a heat transfer law~\cite{GWWC2013,HR2016,J2017,MXDS2018,CMG2017,AP2020,MZCSD2020,ZH2020,GAGMRCH2020,G2021,GARMH2017,GAMRC2018,GASRRH2018,YH2025}.
Yet, in the original derivation of the CA efficiency~\cite{CA1975}, no such symmetric dissipation assumption is required;
Eq.~(\ref{eq.CA}) holds for any asymmetric dissipation under endoreversibility~\cite{ZH2020}.
While this apparent inconsistency at quadratic order was addressed and a general condition was obtained in terms of a nonlinear constitutive relation~\cite{ST2015,ST2015_2}, 
the relationship between the endoreversible model and the LD model is still under discussion~\cite{J2017,ZH2020,GARMH2017,GAMRC2018}.

Here, we derive a general condition under which the CA efficiency is recovered in the quadratic order of  $\eta^*$.
By extending the LD model~\cite{EKLV2010} to include a nonlinear response in the inverse operation time, 
we show that such a nonlinear response generally affects the quadratic order of $\eta^*$.
The CA model and symmetric dissipation condition extended to the nonlinear regime emerge as distinct special cases of the same general condition.
Our condition is also applied to an endoreversible model with a general heat-transfer law, which yields the condition on the heat-transfer coefficients required to recover the CA efficiency.
As the present result is based on a general argument independent of specific models, heat-transfer laws, and scales, 
it provides a broadly applicable result for finite-time heat engines beyond the low-dissipation regime.

Consider a heat engine operating in a finite-time Carnot cycle between hot and cold heat baths at temperatures $T_h$ and $T_c$.
We assume that the heat $Q_i$ ($i=h, c$) absorbed from the heat bath admits the following asymptotic expansion in the inverse duration of $1/\tau_i$ of the corresponding isothermal process, up to quadratic order:
\begin{align}
Q_i=T_i\Delta S_i-\frac{T_i\Sigma_i}{\tau_i}+\frac{T_i\Lambda_i}{\tau_i^2}+\cdots.\label{eq.Q}
\end{align}
Here, $\Delta S_i$ denotes the reversible entropy change during the isothermal process, and $\Sigma_i>0$
and $\Lambda_i$ denote the first- and second-order asymptotic expansion coefficients, respectively.
The work per cycle $W\equiv Q_h+Q_c$ is given by
\begin{align}
W=(T_h-T_c)\Delta S-\frac{T_h\Sigma_h}{\tau_h}-\frac{T_c\Sigma_c}{\tau_c}+\frac{T_h\Lambda_h}{\tau_h^2}+\frac{T_c\Lambda_c}{\tau_c^2},\label{eq.def_W}
\end{align}
where we used $\Delta S_h=-\Delta S_c \equiv \Delta S$. 

The efficiency and power of the cycle are defined by
\begin{align}
&\eta\equiv \frac{W}{Q_h},\label{eq.effi_def}\\
&P\equiv \frac{W}{\tau_h+\tau_c},\label{eq.pow_def}
\end{align}
respectively, where the durations of the adiabatic processes have been neglected~\cite{EKLV2010,C1985}.
Note that in the quasistatic limit $\tau_h \to \infty$ and $\tau_c \to \infty$, the Carnot efficiency $\eta=\eta_{\rm C}$ is recovered, while the power vanishes, $P=0$.

By substituting Eq.~(\ref{eq.def_W}) into Eq.~(\ref{eq.pow_def}), and maximizing it with respect to both $\tau_h$ and $\tau_c$,
we obtain the following conditions for $\tau_h$ and $\tau_c$ at the maximum power:
\begin{align}
&\Sigma_h u_h^2-2\Lambda_hu_h^3=(1-\eta_{\rm C})(\Sigma_c u_c^2-2\Lambda_cu_c^3),\label{eq.cond_Pmax_1}\\
&\eta_{\rm C}\Delta S=2\Sigma_h u_h-3\Lambda_hu_h^2+(1-\eta_{\rm C})(2\Sigma_cu_c-3\Lambda_cu_c^2),\label{eq.cond_Pmax_2}
\end{align}
where we defined $u_h \equiv 1/\tau_h$ and $u_c\equiv 1/\tau_c$.
By substituting the expansion $u_i=a_i^{(1)}\eta_{\rm C}+a_i^{(2)}\eta_{\rm C}^2+\cdots$ into Eqs.~(\ref{eq.cond_Pmax_1}) and (\ref{eq.cond_Pmax_2}), and determining the unknown coefficients $a_i^{(1)}$ and $a_i^{(2)}$, we arrive at the following formula for the efficiency at maximum power (see the detailed derivation of Eqs.~(\ref{eq.cond_Pmax_1})--(\ref{eq.effi_main}) in Appendix~\ref{appendix}):
\begin{align}
\eta^*=\frac{\eta_{\mathrm C}}{2}+\Biggl[\frac{2}{1+q}-\frac{\alpha_h+\alpha_c q^2}{(1+q)^2}\Biggr]\frac{\eta_{\mathrm C}^2}{8}+O(\eta_{\mathrm C}^3),\label{eq.effi_main}
\end{align}
where we defined
\begin{align}
q\equiv \sqrt{\frac{\Sigma_c}{\Sigma_h}}, \quad \alpha_h\equiv \frac{\Lambda_h \Delta S}{\Sigma_h^2}, \quad \alpha_c\equiv \frac{\Lambda_c \Delta S}{\Sigma_c^2}.\label{eq.coefficient}
\end{align}
Because $O(\tau_i^{-3})$ terms and higher-order ones for $Q_i$ do not contribute to $O(\eta_{\mathrm C}^2)$ of $\eta^*$, Eq.~(\ref{eq.effi_main}) is the most general expression within the asymptotic expansion in Eq.~(\ref{eq.Q}).
Note that even if $\Sigma_i$ and $\Lambda_i$ depend on $T_i$, Eq.~(\ref{eq.effi_main}) remains valid when these coefficients are evaluated at the equilibrium reference temperature  $T\approx T_h \approx T_c$, i.e., $\Sigma_i(T_i) \approx \Sigma_i(T)$ and $\Lambda_i(T_i) \approx \Lambda_i(T)$ (see Appendix~\ref{appendix}).

Equation~(\ref{eq.effi_main}) yields the following general condition for the quadratic coefficient to take the CA value $1/8$:
\begin{align}
q^2(1+\alpha_c)=1-\alpha_h.\label{eq.CA_cond}
\end{align}
Results equivalent to Eqs.~(\ref{eq.effi_main}) and (\ref{eq.CA_cond}), written in a different form, 
were previously obtained in terms of a nonlinear constitutive relation~\cite{ST2015,ST2015_2}.
Here, they are presented in a form that explicitly includes the asymptotic expansion coefficients of the extended LD model.

We have two important cases that satisfy the general condition (\ref{eq.CA_cond}).\\
(i) Generalized symmetric dissipation condition:
\begin{align}
\Sigma_h=\Sigma_c, \ \Lambda_h=-\Lambda_c,\label{eq.cond_sym}
\end{align} 
which imposes symmetry on the first-order expansion coefficients $\Sigma_i$ and antisymmetry on the second-order expansion coefficients $\Lambda_i$.
This leads to $q=1$ and $\alpha_h=-\alpha_c$, and Eq.~(\ref{eq.CA_cond}) is thus satisfied.
This generalizes the symmetric dissipation condition in the LD model: 
By putting $\Lambda_h=\Lambda_c=0$, i.e., $\alpha_h=\alpha_c=0$ in Eq.~(\ref{eq.effi_main}), we recover the result in the LD model (Eq.~(10) of Ref.~\cite{EKLV2010}):
\begin{align}
\eta^*=\frac{\eta_{\mathrm C}}{2}+\frac{\eta_{\mathrm C}^2}{4(1+q)}+O(\eta_{\mathrm C}^3),
\end{align}
where  the CA efficiency is achieved under the symmetric condition $q=1$ ($\Sigma_h=\Sigma_c$).
\\
(ii) CA model~\cite{CA1975}:
Under the endoreversible assumption, the working substance has a well-defined temperature.
This assumption implies that no entropy is produced inside the working substance; entropy production arises only from heat transfer across the finite temperature difference between the working substance and the corresponding heat bath.
The heat during an isothermal process is given by $Q_i=T_i^w\Delta S_i$, where $T_i^w$ is the constant temperature of the working substance, 
which ensures that the cycle closes as $\sum_i Q_i/T_i^w=\sum_i \Delta S_i=0$.
By further assuming that the heat flows between the heat bath and the working substance according to Newton's law of cooling 
\begin{align}
Q_i=K_i (T_i-T_i^w)\tau_i, \label{eq.Newton}
\end{align}
where $K_i$ denotes the thermal conductance, we can determine $T_i^w$ and obtain
\begin{align}
&Q_i=\frac{T_i\Delta S_i}{1+\frac{\Delta S_i}{K_i \tau_i}}=T_i\Delta S_i+T_i \sum_{n=1}^\infty (-1)^n \frac{\Delta S_i^{n+1}}{K_i^n\tau_i^n},
\end{align}
where $|\Delta S_i/(K_i\tau_i)| < 1$ is assumed.
This identifies the first- and second-order asymptotic expansion coefficients as
\begin{align}
\Sigma_i=\frac{\Delta S_i^2}{K_i}, \ \Lambda_i=\frac{\Delta S_i^3}{K_i^2},
\end{align}
where the expansion coefficients at different orders are correlated as a consequence of endoreversibility and the well-defined temperature of the working substance.
This gives $q=\sqrt{K_h/K_c}$ and $\alpha_h=-\alpha_c=1$, and Eq.~(\ref{eq.CA_cond}) is thus satisfied even in the asymmetric case ($q \ne 1$).

If endoreversibility is retained but Newton's law of cooling is replaced by a different heat-transfer law, the CA efficiency is no longer guaranteed, and an additional condition on the heat-transfer coefficients may be required.
To illustrate this point, we consider the linear irreversible heat-transfer law~\cite{CY1989}
\begin{align}
Q_i=K_i \left(\frac{1}{T_i^w}-\frac{1}{T_i}\right)\tau_i.\label{eq.inverse}
\end{align}
An analytical expression for $\eta^*$ is available for this model:
\begin{eqnarray}
\eta^*=\frac{\eta_{\rm C}}{2-\frac{\eta_{\rm C}}{1+\sqrt{\frac{K_c}{K_h}}}}=\frac{\eta_{\rm C}}{2}+\frac{\eta_{\rm C}^2}{4(1+\sqrt{K_c/K_h})}+O(\eta_{\rm C}^3).\label{eq.eta_Pmax_inverse}\nonumber\\
\end{eqnarray}
We can recover Eq.~(\ref{eq.eta_Pmax_inverse}) up to the quadratic order from Eqs.~(\ref{eq.effi_main}) and (\ref{eq.coefficient}) with the asymptotic expansion coefficients:
\begin{align}
\Sigma_i=\frac{T_i^2\Delta S_i^2}{K_i}, \ \Lambda_i=\frac{2T_i^4\Delta S_i^3}{K_i^2}.
\end{align}
This yields $q=\sqrt{K_h/K_c}$ and $\alpha_h=-\alpha_c=2$.
In contrast to the CA model~\cite{CA1975}, the symmetric condition $K_h=K_c$, which leads to the generalized symmetry in Eq.~(\ref{eq.cond_sym}), is necessary to satisfy Eq.~(\ref{eq.CA_cond}) and recover the CA coefficient $1/8$.

A natural question is why Newton's law of cooling (\ref{eq.Newton}) always yields the CA efficiency while the linear irreversible law (\ref{eq.inverse}) requires the symmetric condition.
To answer this question, we consider the following general heat-transfer law:
\begin{align}
Q_i=\kappa_{i1}(T_i)(T_i-T_i^w)\tau_i+\kappa_{i2}(T_i)(T_i-T_i^w)^2\tau_i+\cdots,
\end{align}
which includes nonlinear response to the temperature difference.
Newton's law of cooling ($\kappa_{i1}=K_i$ and $\kappa_{i2}=0$) and the linear irreversible law ($\kappa_{i1}=K_i/T_i^2$ and $\kappa_{i2}=K_i/T_i^3$) are special cases of this generalized law.
The asymptotic expansion coefficients are identified as
\begin{align}
\Sigma_i=\frac{\Delta S_i^2}{\kappa_{i1}}, \quad \Lambda_i=\frac{\Delta S_i^3}{\kappa_{i1}^2}\left(1+\frac{T_i\kappa_{i2}}{\kappa_{i1}}\right).
\end{align}
For this generalized heat-transfer law, the condition~(\ref{eq.CA_cond}) for recovering the CA efficiency becomes
\begin{align}
\frac{\kappa_{h2}}{\kappa_{h1}^2}=\frac{\kappa_{c2}}{\kappa_{c1}^2}.\label{eq.CA_cond_gen}
\end{align}
Equation~(\ref{eq.CA_cond_gen}) shows that the absence of the nonlinear response, $\kappa_{i2}=0$, automatically satisfies the condition, as in Newton's law of cooling (\ref{eq.Newton}).
For the linear irreversible law (\ref{eq.inverse}), by contrast, Eq.~(\ref{eq.CA_cond_gen}) reduces to $K_h=K_c$, corresponding to the generalized symmetric condition (\ref{eq.cond_sym}).
Because the linear irreversible law possesses the nonlinear response in the temperature difference, it requires the symmetry.

In conclusion, we derived the general condition, Eq.~(\ref{eq.CA_cond}), under which the CA efficiency is recovered through quadratic order beyond the low-dissipation regime. 
The generalized symmetric dissipation condition and CA model are shown to be distinct special cases satisfying the same general condition.
While the LD model appears to contradict the CA result~\cite{CA1975}, which allows asymmetric dissipation, this apparent discrepancy is resolved by explicitly accounting for nonlinear-response terms.
This therefore clarifies the scope of applicability of the standard LD model~\cite{EKLV2010}.
The present result is genuinely a nonlinear nonequilibrium result, which clarifies the nonlinear origin of the universality of the CA efficiency.

The asymptotic expansion coefficients $\Sigma_i$ and $\Lambda_i$ in Eq.~(\ref{eq.Q}) generally depend on the protocol used to operate the heat engine over a given duration $\tau_i$.
It has been suggested that dissipation minimization over the operating protocol can give rise to symmetry in the coefficients $\Sigma_i$~\cite{AP2020}.
Whether the same mechanism also constrains the coefficients $\Lambda_i$ in an analogous manner remains an interesting open question.
Besides, how the introduction of nonlinear response affects the trade-off relation between power and efficiency~\cite{RSP2016,SST2016,BS2020,ZCMSD2023,DS2018,VS2024} and Pareto optimality among multiple performance criteria~\cite{AS2021,ERAPLN2023,I2024,AMRM2026} deserves further investigation.

\acknowledgments
The author used ChatGPT (OpenAI, GPT-5.6 Sol) to assist with mathematical derivations and manuscript editing.

\appendix
\section{Derivation of Eqs.~(\ref{eq.cond_Pmax_1})--(\ref{eq.effi_main})}\label{appendix}
By substituting Eq.~(\ref{eq.def_W}) into Eq.~(\ref{eq.pow_def}), and maximizing it with respect to both $\tau_h$ and $\tau_c$, we obtain the following conditions at the maximum power:
\begin{align}
&\frac{\partial P}{\partial \tau_h}=\frac{\frac{\partial W}{\partial \tau_h}(\tau_h+\tau_c)-W}{(\tau_h+\tau_c)^2}=0,\\
&\frac{\partial P}{\partial \tau_c}=\frac{\frac{\partial W}{\partial \tau_c}(\tau_h+\tau_c)-W}{(\tau_h+\tau_c)^2}=0,
\end{align}
which are equivalent to
\begin{align}
&\frac{\partial W}{\partial \tau_h}=\frac{W}{\tau_h+\tau_c},\label{eq.maximize_W1}\\
&\frac{\partial W}{\partial \tau_c}=\frac{W}{\tau_h+\tau_c}.\label{eq.maximize_W2}
\end{align}
Thus, we obtain
\begin{align}
\frac{\partial W}{\partial \tau_h}=\frac{\partial W}{\partial \tau_c}.\label{eq.maximize_W3}
\end{align}
By substituting Eq.~(\ref{eq.def_W}) into both sides of Eq.~(\ref{eq.maximize_W3}), we derive
\begin{align}
\frac{T_h\Sigma_h}{\tau_h^2}-\frac{2T_h\Lambda_h}{\tau_h^3}=\frac{T_c\Sigma_c}{\tau_c^2}-\frac{2T_c\Lambda_c}{\tau_c^3},
\end{align}
which reduces to Eq.~(\ref{eq.cond_Pmax_1}) by introducing $u_i=1/\tau_i$ and dividing both sides by $T_h$.

By multiplying both sides of Eq.~(\ref{eq.maximize_W1}) by $\tau_h$ and both sides of Eq.~(\ref{eq.maximize_W2}) by $\tau_c$, and summing them, we obtain
\begin{align}
\frac{\partial W}{\partial \tau_h}\tau_h+\frac{\partial W}{\partial \tau_c}\tau_c=W.\label{eq.maximize_W4}
\end{align}
By substituting Eq.~(\ref{eq.def_W}) into both sides of Eq.~(\ref{eq.maximize_W4}), we obtain
\begin{align}
(T_h-T_c)\Delta S=\frac{2T_h\Sigma_h}{\tau_h}+\frac{2T_c\Sigma_c}{\tau_c}-\frac{3T_h\Lambda_h}{\tau_h^2}-\frac{3T_c\Lambda_c}{\tau_c^2},
\end{align}
which reduces to Eq.~(\ref{eq.cond_Pmax_2}) by introducing $u_i=1/\tau_i$ and dividing both sides by $T_h$.

By substituting the expansion $u_i=1/\tau_i=a_i^{(1)}\eta_{\rm C}+a_i^{(2)}\eta_{\rm C}^2+\cdots$ into Eqs.~(\ref{eq.cond_Pmax_1}) and (\ref{eq.cond_Pmax_2}), we can determine the coefficients $a_i^{(1)}$ and $a_i^{(2)}$ order by order as
\begin{align}
&a_h^{(1)}=\frac{\Delta S}{2\Sigma_h(1+q)},\label{eq.expand_ah1}\\
&a_c^{(1)}=\frac{\Delta S}{2q\Sigma_h(1+q)},\label{eq.expand_ac1}\\
&a_h^{(2)}=\frac{\Delta S}{8\Sigma_h(1+q)^3}[\alpha_cq^2+2\alpha_hq+3\alpha_h+2q^2+2q],\label{eq.expand_ah2}\\
&a_c^{(2)}=\frac{\Delta S}{8q\Sigma_h(1+q)^3}[3\alpha_cq^2+2\alpha_cq+\alpha_h+4q^2+6q+2].\label{eq.expand_ac2}
\end{align}
Here, we also note the following relations:
\begin{align}
&\Sigma_h a_h^{(1)}+\Sigma_c a_c^{(1)}=\frac{\Delta S}{2}\label{eq.relation1},\\
&\Sigma_h a_h^{(2)}+\Sigma_c a_c^{(2)}=\frac{3}{2}\Lambda_h (a_h^{(1)})^2+\frac{3}{2}\Lambda_c (a_c^{(1)})^2+\Sigma_c a_c^{(1)},\label{eq.relation2}
\end{align}
where the former readily follows from Eqs.~(\ref{eq.expand_ah1}) and (\ref{eq.expand_ac1}) and the latter is obtained from $O(\eta_{\rm C}^2)$ of Eq.~(\ref{eq.cond_Pmax_2}).
By substituting $u_i=a_i^{(1)}\eta_{\rm C}+a_i^{(2)}\eta_{\rm C}^2+\cdots$ 
into Eqs.~(\ref{eq.Q}) and (\ref{eq.def_W}), we obtain
\begin{align}
Q_h
&=T_h\bigl[\Delta S-\Sigma_h a_h^{(1)}\eta_{\rm C}\bigr]
 +O(\eta_{\rm C}^2),
\label{eq.Q*}\\
W
&=T_h\bigl[\Delta S-\Sigma_h a_h^{(1)}
 -\Sigma_c a_c^{(1)}\bigr]\eta_{\rm C}
\nonumber\\
&\quad+T_h\bigl[
 -\Sigma_h a_h^{(2)}-\Sigma_c a_c^{(2)}
 +\Sigma_c a_c^{(1)}
\nonumber\\
&\qquad
 +\Lambda_h(a_h^{(1)})^2
 +\Lambda_c(a_c^{(1)})^2
 \bigr]\eta_{\rm C}^2
 +O(\eta_{\rm C}^3).
\label{eq.W*}
\end{align}
By using Eqs.~(\ref{eq.relation1}) and (\ref{eq.relation2}) in Eq.~(\ref{eq.W*}), we obtain
\begin{align}
W=T_h\frac{\Delta S}{2}\eta_{\rm C}-\frac{T_h}{2}\left[\Lambda_h(a_h^{(1)})^2+\Lambda_c(a_c^{(1)})^2\right]\eta_{\rm C}^2+O\left(\eta_{\rm C}^3\right).\label{eq.W*_2}
\end{align}
By substituting Eqs.~(\ref{eq.Q*}) and (\ref{eq.W*_2}) into Eq.~(\ref{eq.effi_def}), and using Eqs.~(\ref{eq.expand_ah1}) and (\ref{eq.expand_ac1}), we derive Eq.~(\ref{eq.effi_main}).

Subsequently, we show that Eq.~(\ref{eq.effi_main}) remains valid even when $\Sigma_i$ and $\Lambda_i$ depend on $T_i$.
We write $T_h\approx T(1+\gamma \eta_{\rm C})$ and $T_c\approx T[1-(1-\gamma)\eta_{\rm C}]$ by neglecting $O(\eta_{\rm C}^2)$,
where $\gamma$ ($0\le \gamma \le 1$) denotes a parameter specifying how the temperature difference is distributed around the equilibrium reference temperature $T$.
We then expand $\Sigma_i(T_i)$ around $T$ as
\begin{align}
\Sigma_i(T_i)=\Sigma_i+s_i\eta_{\rm C}+O(\eta_{\rm C}^2),
\end{align}
where we write $\Sigma_i(T)=\Sigma_i$ for simplicity and the first-order expansion coefficients $s_i$ are defined as
\begin{align}
&s_h\equiv \gamma T \Sigma_h'(T),\\
&s_c\equiv -(1-\gamma)T\Sigma_c'(T),
\end{align}
where the prime denotes differentiation.
With this temperature dependence included, Eq.~(\ref{eq.relation1}) remains unchanged, whereas Eq.~(\ref{eq.relation2}) becomes
\begin{align}
\begin{aligned}
\Sigma_ha_h^{(2)}
+\Sigma_ca_c^{(2)}
={}&
\frac32\Lambda_h(a_h^{(1)})^2
+\frac32\Lambda_c(a_c^{(1)})^2
\\
&+\Sigma_ca_c^{(1)}
-s_ha_h^{(1)}
-s_ca_c^{(1)}.
\end{aligned}\label{eq.relation3}
\end{align}
The work (\ref{eq.W*}) also changes as
\begin{align}
\begin{aligned}
W
={}&
T_h\left[
\Delta S
-\Sigma_ha_h^{(1)}
-\Sigma_ca_c^{(1)}
\right]\eta_{\rm C}
\\
&+
T_h\Big[
-\Sigma_ha_h^{(2)}
-\Sigma_ca_c^{(2)}
+\Sigma_ca_c^{(1)}
\\
&\qquad
+\Lambda_h(a_h^{(1)})^2
+\Lambda_c(a_c^{(1)})^2
\\
&\qquad
-s_ha_h^{(1)}
-s_ca_c^{(1)}
\Big]\eta_{\rm C}^2
+O(\eta_{\rm C}^3).
\end{aligned}\label{eq.W*_3}
\end{align}
Here, we approximate $\Lambda_i(T_i)\approx \Lambda_i(T)$ and just write $\Lambda_i(T)=\Lambda_i$ in Eqs.~(\ref{eq.relation3}) and (\ref{eq.W*_3}) because the temperature dependence of $\Lambda_i(T_i)$ does not contribute at this order.
By substituting Eqs.~(\ref{eq.relation1}) and (\ref{eq.relation3}) into Eq.~(\ref{eq.W*_3}), Eq.~(\ref{eq.W*_2}) is recovered because the $s_i$-dependent terms cancel exactly.
We thus conclude that Eq.~(\ref{eq.effi_main}) remains valid, provided that the coefficients appearing in Eq.~(\ref{eq.effi_main}) are evaluated at the equilibrium reference temperature $T$.

\end{document}